\documentclass{PoS}
\title{The supernova remnant population in the very--high--energy sky: prospects for CTA}

\ShortTitle{Supernova remnants in the very--high--energy sky}

\author{\speaker{Pierre Cristofari}\\
        Columbia University\\
        E-mail: \email{pc2781@columbia.edu}}

\abstract{The detection of very--high--energy gamma rays from supernova remnant shells testifies of the acceleration of particles at strong shocks. Many aspects of the particle acceleration remain however unclear. The study of individual objects is very helpful, but the study of the entire  population of SNRs detected in this range and its characteristics can also bring valuable science.
Using Monte--Carlo simulations, the population of shells bright in the TeV and multi--TeV range can be simulated. The results of these simulations aim at being compared with observations of instruments operating in these ranges, such as the Cherenkov Telescope Array (CTA). Our results suggest that CTA should be able to effectively constrain the slope of particles accelerated at SNRs and the electron--to--proton ratio.
}

\FullConference{The European Physical Society Conference on High Energy Physics\\
		5-12 July, 2017\\
		Venice}

\begin{document}

\section{Introduction}
The Galactic population of supernova remnants (SNRs) in commonly thought to play an important part in the acceleration of Galactic cosmic rays (CRs).  A detailed description and the subtleties of this contribution remain however missing. The detection of many SNRs in the TeV range by current instruments~\cite{donath2016} has considerably improved the  understanding of the community, but many of the parameters describing the acceleration of very--high--energy particles, such as for example  the slope of the particles accelerated at the shock, the maximum energy of particles, the cosmic--ray efficiency, the magnetic field amplification,  remain unknown. Even more than the individual study of SNRs, the study of the population of SNRs as such could drastically constrain the parameters governing particle acceleration at SNR shocks.

The Cherenkov Telescope Array (CTA) will perform a Galactic survey with an unprecedented sensitivity  in the TeV and multi--TeV range, thus expected to increase the population of SNRs in gamma rays. Two strategies have been especially proposed: a Galactic Plane Survey where the typical sensitivity above 1 TeV can reach $\approx 1$~mCrab, and an all--sky survey where the sensitivity above 1 TeV can reach $\approx 3$~mCrab~\cite{CTA2013}.

We propose to simulate the population of Galactic SNRs detectable in gamma rays by CTA. These simulations aim at being confronted with future observations, thus providing material to improve our understanding of particle acceleration at SNR shocks.  

\section{Method}
We rely on Monte Carlo methods to simulate the population of SNRs potentially detectable by CTA.  
The time and location of supernovae in the Galaxy are simulated, assuming a rate of 3 SN per century, and a spatial distribution described as in~\cite{faucher}. Four types of progenitors are considered:~thermonuclear (type Ia) and core--collapse (types Ib/c, IIP, IIb). The relative rates and typical parameters associated to each type, such as the total supernova explosion energy, the velocity of the wind and the mass of the ejecta, are adopted as in~\cite{seo}, so that every simulated supernova is assigned a type and corresponding parameters. 
At the location of each supernova, the typical value of the interstellar medium (ISM) is derived from surveys of atomic and molecular hydrogen~\cite{H1,H2}. The evolution of the shock radius $R_{\rm sh}$ and velocity $u_{\rm sh}$ is computed using the analytical and semi--analytical descriptions of~\cite{chevalier,pz05}. 

We then compute the gamma--ray luminosity of each SNR.  Both contributions of protons and electrons are taken into account. At the shock, the particles are assumed to be accelerated with a slope following a power--law in momentum $f(p) \propto p^{-\alpha}$, where $\alpha$ is treated as a parameter in the range $4.1 - 4.4$. At the shock, we assumed that a fraction $\xi_{\rm CR}$ of the ram pressure of the shock expanding through the ISM is converted into CRs, where $\xi_{\rm CR}$ is typically $\approx 0.1$, and the shock compression factor is $\sigma=4$.  
Following~\cite{pz03,pz05}, we solve the transport equation inside the SNR, thus deriving the distribution of CRs, and we solve the gas continuity equation, thus deriving the structure of the interior of the SNR. 
The hadronic contribution to the gamma--ray spectrum is then calculated following the approach of~\cite{kelner2006}, weighted by a factor 1.8 to take into account nuclei heavier than hydrogen. 
The spectrum of electrons is parametrized at low energies adopting the same spectral shape as protons $\propto p^{-\alpha}$, weighted by a factor $K_{\rm ep}$, and the gamma--ray luminosity from inverse Compton scattering of electrons on the cosmic microwave background is computed following the description proposed by~\cite{gould}.

This approach was used in~\cite{cristofari2013} to propose a test of the SNR paradigm for the origin of Galactic CRs and in~\cite{cristofari2017} to investigate prospects for CTA. 

\section{Results}

\begin{figure}
\centering
\includegraphics[width=.5\textwidth]{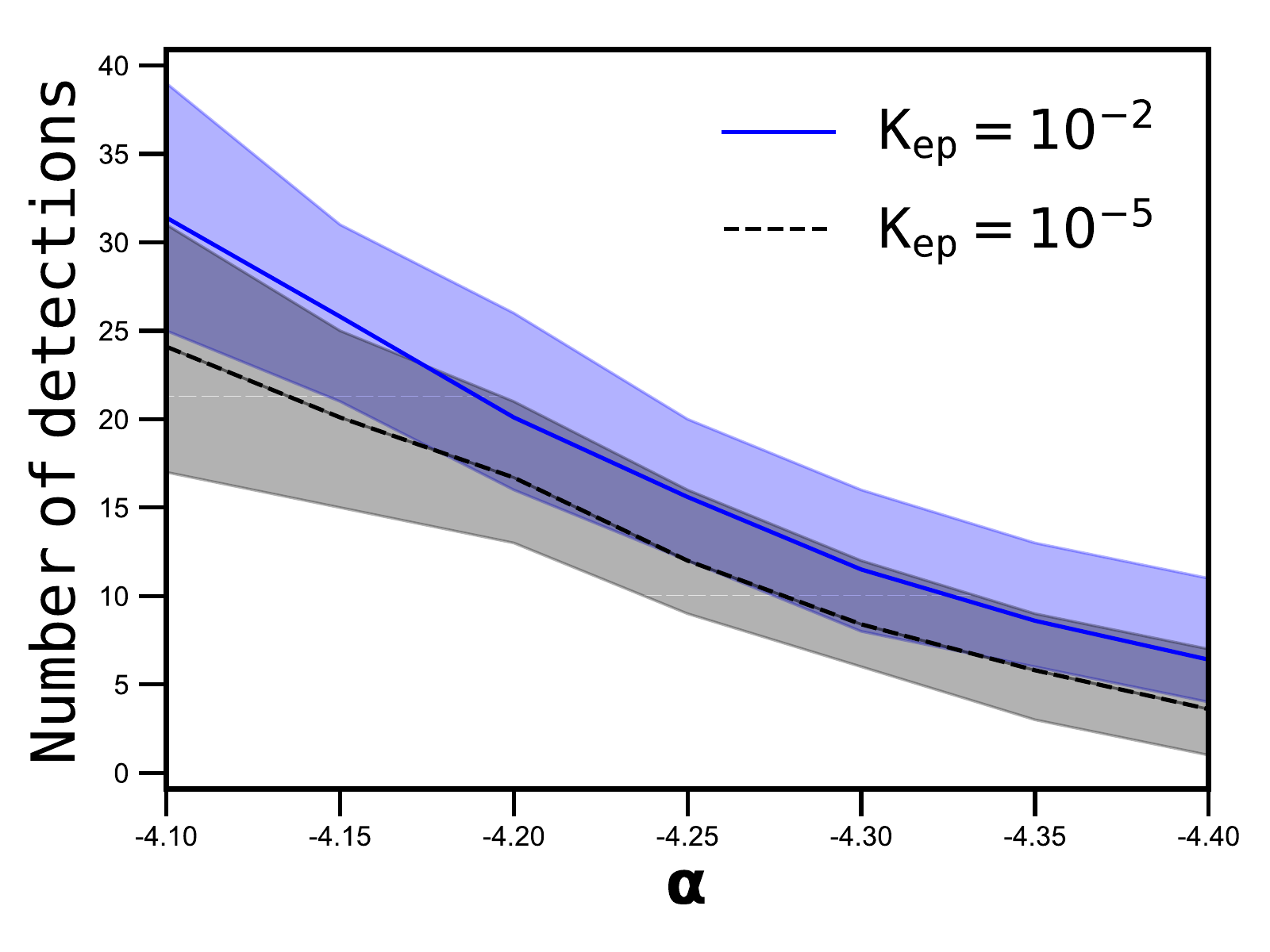}
\caption{SNRs in the simulated Galactic plane survey of CTA with integral gamma--ray flux F($>$~10 TeV)~$\geq$~1~mCrab, as a function of the parameter $\alpha$. The blue (solid) and black (dashed) curve correspond respectively to $K_{\rm ep }= 10^{-2}$ and $K_{\rm ep }= 10^{-5}$. In each case the +/- standard deviation is shown.}
\label{fig:alphaplot}
\end{figure}

We investigate the effects of the main parameters used to describe particle acceleration at SNR shocks. 
Fig.~\ref{fig:alphaplot} shows the number of SNRs with integral gamma--ray fluxes F($>$~10 TeV) greater than ~10~mCrab, the typical sensitivity of CTA in the planned GPS survey. The mean number of detections varies between $\approx 30^{+8}_{-7}$ and $\approx 4^{+2}_{-2}$ for the two extreme situations: $\alpha = 4.1 \; , \;  K_{\rm ep}=10^{-2}$ and $\alpha = 4.4 \; ,\;K_{\rm ep}=10^{-5}$, respectively. The plot suggests that the number of shells detected during the CTA survey should be able to constrain the slope of accelerated particles $\alpha$, but can not strongly constrain $K_{\rm ep}$.
 Results for fluxes integrated above 1~TeV are presented in~\cite{cristofariICRC}, and suggest that the TeV range is on the other hand more helpful than the 10 TeV range to constrain $K_{\rm ep}$. 
 A more detailed study can be found in~\cite{cristofari2017}. 
 
 \section{Conclusions}
 The CTA telescope will perform a systematic survey of the entire sky with an unequaled sensitivity in the TeV and multi--TeV range. It is therefore a natural tool for the investigation of populations of objects, and especially SNRs. Simulations of the SNR population highlight the role played by the different acceleration parameters discussed above, and show that the different plausible values for these parameters can lead to remarkably different populations. 
The GPS performed by H.E.S.S. lead to the detection of 78 sources~\cite{donath2016}, but only 8 of them are clearly identified as SNR shells. The large number of unidentified sources ($\approx 47$) did not provide yet sufficient information to firmly constrain $\alpha$ or $K_{\rm ep}$. 
 They therefore strongly suggest that CTA will be able to discriminate between the different simulated scenarios, thus providing a new tool to constrain the parameters describing particle acceleration.

 \section*{Acknowledgements}
This work was conducted in the context of the CTA Consortium. The author gratefully acknowledges financial support from the agencies and organizations listed here: http://www.cta-observatory.org/consortium\_acknowledgments, and thank the CTA consortium. PC acknowledges support from the Columbia University Frontiers of Science fellowship.

\end{document}